\documentclass[pre,aps,twocolumn]{revtex4-1}
\usepackage{amsfonts}
\usepackage[latin1]{inputenc}
\usepackage[english]{babel}
\usepackage{graphicx}
\usepackage{color}
\usepackage{amsmath}
\begin{document}

\title{\bf\Large Two--step condensation of the charged Bose gas}

\author{R. L. Delgado, P. Bargue\~no$^{*}$ and F. Sols}
\affiliation{Departamento de F\'{\i}sica de Materiales,
Universidad Complutense de Madrid, {\it E-28040},
Madrid, Spain ($^*$p.bargueno@fis.ucm.es)
}

\begin{abstract}
The condensation of the spinless ideal charged Bose gas in the presence of a
magnetic field is revisited. The conventional approach is extended to include the macroscopic occupation
of excited kinetic states lying in the lowest Landau level, which plays an essential role in the case of large magnetic fields.
In that limit, signatures of two diffuse phase transitions (crossovers) appear in the specific heat. In particular, at temperatures lower than the cyclotron frequency, the system behaves as an effectively one-dimensional free boson system, with the specific heat equal to $\frac{1}{2}Nk_{B}$ and a gradual condensation at lower temperatures.
\end{abstract}
\maketitle

\section{Introduction}

The equilibrium behavior of a charged Bose gas (CBG) in the presence of a magnetic field is a fundamental problem in quantum statistical mechanics. It was first addressed by Schafroth \cite{Schafroth} at a time where the BCS theory of superconductivity had not yet been discovered. Although it was soon recognized that a model of non-interacting bosons had little predictive power for superconductivity, the CBG problem has continued to attract considerable attention because of its appealing simplicity and as a model to simulate astrophysical scenarios where charged particles are subject to extremely large magnetic fields, or neutral particles move in a fast rotating background. 
It is also interesting as a preliminary step in the study of the condensation of interacting bosons in a magnetic field or under rotation.

For the non--interacting case, Schafroth noticed that, in the presence of a magnetic field, Bose--Einstein condensation (BEC) no longer occurs 
in a strict sense, since at low enough temperatures, the system behaves as one-dimensional (1D) and a non-interacting 1D boson system is known to exhibit not a sharp but a gradual (diffuse) BEC transition. Later, May \cite{May1,May2} investigated the condensation
in the non--relativistic case and showed that a sharp transition occurs for dimension $\ge 5$. Daicic and
coworkers \cite{Daicic1,Daicic2} extended May's findings to the relativistic high-temperature case.
Using a different definition of BEC, Toms \cite{Toms1,Toms2,Toms3} proved that BEC in the presence of a
uniform magnetic field does not
occur in any number of spatial dimensions and Elmfors and coworkers \cite{Elmfors} considered that, in
the three--dimensional (3D) case, although a true condensate is not formed, the Landau ground state can accommodate
a large boson number. In this spirit, Rojas \cite{Rojas1,Rojas2} found that BEC may occur in the presence of a homogeneous
magnetic field, but there is no critical temperature at which condensation starts, the phase
transition being diffuse.
Diffuse phase transitions are those not having a definite critical temperature, but an interval of
temperatures along which the transition occurs gradually. The concept was already introduced in the study of phase transitions which occur in certain ferroelectric
materials \cite{Smolenski1954}.  In this paper we treat the terms {\it diffuse transition} and {\it crossover} as synonyms.
The notion of BEC of a CBG in the presence of a magnetic field in 3D continued to be studied by Toms and coworkers \cite{Toms4,Toms5}, who showed that, although
there is no BEC in the sense of a sharp phase transition, the specific heat exhibits a clear maximum that can be used to define a critical temperature. Using that definition, the authors inferred that
the critical temperature increases with the magnetic field \cite{Toms5}, reaching the usual value for the
3D free Bose gas when no magnetic field is present.
The extension of these studies to the case of vector bosons in strong magnetic fields and at high
temperatures was performed by Khalilov and coworkers  \cite{Khalikov1997,K98,K99}. Specifically,
the condensation and effective magnetization of a charged vector boson gas were studied, showing that there is no true BEC as
well, although a significant amount of bosons can accumulate in the ground state at low temperatures.
More recently, the magnetic properties of charged spin--1 Bose gases in an external magnetic field have been revisited, with focus on the competition between diamagnetism and paramagnetism \cite{Jian2011}.

In the particular case of a large magnetic field, the main feature of a CBG is the coexistence of highly degenerate Landau levels and a fine structure of levels with nonzero momentum parallel to the magnetic field. We propose that a CBG under a large magnetic field can also be used as a crude model to understand basic features of systems where a fine sublevel structure coexists with sectors characterized by highly discrete quantum numbers. Specifically, we have in mind the condensation of bosonic molecules, whose level structure is hierarchical, with their rotational states structured into vibrational sectors which in turn can be grouped into largely spaced electronic levels.

Thus we are motivated by the study of molecular BECs, where a complex molecular level structure may give rise to multi-step BEC. Molecular condensates are expected to display a wealth of fundamental phenomena. For instance, it has been proposed that molecular condensates may permit the experimental study of low-energy parity violation \cite{Bargueno2012}. Specifically, the fact that recently both homonuclear \cite{Danzl2010}, heteronuclear \cite{Aikawa2010,Ni2010}
ultracold molecules and heteronuclear molecular ions \cite{Staanum2010,Schneider2010} have been produced in the
rovibrational ground state, suggests that molecular BEC may not lie too far in the future. The conclusion is that the
study of multi-step condensation of the CBG is interesting not only from an academic point of view, but it may provide a qualitative understanding of the condensation of molecules with a hierarchical level structure.

We consider a wide range of magnetic fields because we are interested in understanding the general trends. This means that in our analysis we include magnetic fields so large that would be unrealistic in some contexts \cite{comment1,Kaplan1986,Chakrabarty1997}. Comparing our results with the previous related literature on the CBG (see, for example, Refs. \onlinecite{Toms4,Toms5}), we notice the existence of a term that has so far been neglected. This is the occupation of states in the lowest Landau level but with arbitrary kinetic energy in the direction parallel to the magnetic field. This term is particularly important at large magnetic fields and in fact is responsible for the two-step condensation we refer to in the title and which we shall discuss in detail. We advance here that, as the temperature descends, a first step is defined by the onset of a preferential occupation of states in the lowest Landau level but without any of them absorbing a large fraction of the bosons. Then a plateau is reached where the system behaves as effectively one-dimensional. The second condensation step occurs at even lower temperatures, when the true one-particle ground state becomes host to a macroscopic fraction of the total boson number. By the time this occurs, the systems behaves as one-dimensional, which translates into a diffuse Bose-Einstein condensation, as opposed to the conventional, sharp phase transition which is characteristic of three dimensions. Interestingly, these two distinct, both gradual condensation steps have been noted by Ketterle and coworkers \cite{Ketterle1996,Druten1997} in a different but somewhat analogous context, namely, that of condensation in a strongly anisotropic harmonic trap. As compared to a CBG in a large magnetic field, the role of the Landau levels is played there by the transverse subbands characterizing the motion in the most confined direction. In Refs. \onlinecite{Ketterle1996,Druten1997} the above mentioned term of the occupation of the states with the lowest value of the most discrete quantum number, was correctly included, which resulted in the prediction of a two-step condensation.

\section{Structure of the level population}

\label{conventional}
Let us consider a charged gas of bosonic particles confined in a 3D box of volume $A L$, where $A$ is the
area in the $x$--$y$ plane and $L$ is the length in the $z$ direction.
Away from the edges of the box, the energy of a charged particle in a uniform magnetic
field $B$ pointing in the $z$ direction
is quantized as \cite{Landau}
\begin{equation}
\label{landaulevels}
E_{n}(n_{z})=\hbar\omega \left(n+\frac{1}{2}\right)+\varepsilon n_{z}^{2},
\end{equation}
where  $\omega= q B / m c$ is the cyclotron frequency, $\varepsilon \equiv \hbar^{2} \pi^{2}/2mL^{2}$ characterizes the level spacing in the $z$ direction, and both $n$ and $n_z$ run over natural numbers (we assume hard wall boundary conditions at $z=0$ and $z=L$). The index $n$ is said to characterize the Landau level.

The area $A$ is assumed to be large enough for the role of edge states in the $x-y$ plane to be negligible. Specifically, $l^2\ll A$, where $l^{2}=\hbar c / qB$ is the magnetic length squared.

We start by writing the total number of particles $N$ as ($\hbar=1$)
\begin{equation}
N = d\sum_{n_{z} = 0}^\infty \sum_{n = 0}^\infty \frac{z}{\exp\left( \beta \varepsilon n_{z}^2 +
\beta\omega n\right)-z},
\end{equation}
where $\beta\equiv (k_{B}T)^{-1}$, $d=A/l^2$ is the Landau degeneracy, and $z\equiv \exp(\beta\mu)$ is the fugacity, with $\mu$ the chemical potential [which absorbs the $1/2$ term in (\ref{landaulevels})]. We partition the total particle number into four different groups:
\begin{equation}
  N = N_0 + N_k + N_{m} + N_{mk} \, .
\end{equation}
$N_0$ is the number of particles in the one-particle ground state ($n=n_z=0$),
\begin{equation}\label{term_0}
  N_0 = d\frac{z}{1-z};
\end{equation}
$N_k$ is the number of particles in the ground state of the magnetic level but in excited kinetic levels ($n=0$ and $n_z \neq 0$),
\begin{equation}
N_k = d\sum_{n_{z}= 1}^{\infty} \frac{z}{\exp\left( \beta \varepsilon n_{z}^2\right)-z};
\end{equation}
$N_{m}$ is the number of particles in the kinetic ground state but in excited magnetic states,
\begin{equation}\label{term_v}
  N_{m}=d\sum_{n=1}^{\infty}\frac{z}{\exp\left(\beta\omega n\right)-z};
\end{equation}
and $N_{mk}$ is the number of particles in excited states of both magnetic and kinetic degrees of freedom,
\begin{equation}
  N_{mk}=d\sum_{n=1}^{\infty}\sum_{n_{z}= 1}^{\infty} \frac{z}{\exp\left(\beta\varepsilon n_{z}^2 + \beta\omega n\right)-z}.
\end{equation}
The term $N_k$ can be approximated by
\begin{equation}\label{term_k}
N_k \simeq d \eta_k  g_{1/2}(z).
\end{equation}
where $\eta_k=(\pi /4)^{1/2} (\beta \varepsilon)^{-1/2}$ and $g_{\alpha}(z)\equiv \sum_{j=1}^{\infty}z^{j}/j^{\alpha}$ stands
for the polylogarithm function of order $n$ \cite{Huangbook}. Here, the identity $\lim_{\lambda \rightarrow 0}\sum_{j=1}^\infty\lambda f(\lambda j) = \int_0^\infty f(x)$ has been used assuming that $\lambda=\sqrt{\beta \varepsilon}$ is sufficiently small. Similarly, $N_{m}$ and $N_{mk}$ can be approximated as:

\begin{equation}\label{term_m}
N_{m} \simeq d \eta_m  g_{1}(z)
\end{equation}
and
\begin{equation}\label{term_mk}
N_{mk} \simeq d \eta_m \eta_k g_{3/2}(z),
\end{equation}
where $\eta_m \equiv (\beta \omega)^{-1}$. We can group these approximations to write the total particle number as
\begin{equation}\label{analy_equation}
  \frac{N}{d} = g_{0}(z) + \eta_{k} g_{1/2}(z) + \eta_{m} g_{1}(z)
+ \eta_{k}\eta_{m}g_{3/2}(z).
\end{equation}

As noted by the authors of Ref. \onlinecite{Druten1997} in a similar context (with harmonic instead of hard-wall confinement),
the expression Eq. (\ref{analy_equation}) is an excellent and controlled approximation for the
whole temperature regime. We note that, in this equation, the second, third, and fourth terms become dominant in the respective cases $\eta_{k}\gg1$, $\eta_{m}\gg 1$ and both $\eta_{k},\eta_{m} \gg 1$.

The factor $d$ appearing in all the occupation numbers, Eqs. (\ref{term_0})-(\ref{analy_equation}), reveals the existence of Landau degeneracy, which in particular reflects a multiplicity of one-particle ground states where bosons are in both the kinetic and magnetic ground state ($n=n_z=0$).

The second term, $N_k$, represents the occupation of particles lying in the ground Landau manifold but
in excited kinetic states (with motion in the $z$ direction). We point out that this term has often been neglected in studies of the
CBG. We show below that $N_k$ plays a most important role in the thermodynamic properties of the CBG under large magnetic fields.
The third term, $N_{m}$, which accounts for those bosons which are in the kinetic
ground state but in excited magnetic states, will not be studied here because it
acquires importance only for very small magnetic fields, that is, when the energy of the first magnetic excited
state is much lower than the energy of the first kinetic state ($\omega \ll \varepsilon$).
This would describe a two-dimensional boson gas which will not be considered here.
The last term, $N_{mk}$, corresponds to the conventional term in which the bosons are in both magnetic and
kinetic excited states. It is the dominant term at high enough temperature, where the 3D limit is recovered ($\eta_m,\eta_k \gg 1$).

Finally, we note that the order of the polylogarithm functions appearing in Eq. (\ref{analy_equation}), $g_{d/2}(z)$,
corresponds to the population of excited states of free particles in $d$ dimensions.

\section {Numerical results for $N_{0}$ and $N_{k}$}

For zero magnetic field and vanishing $\varepsilon$ the CBG resembles a 3D
free Bose gas, showing BEC at the usual
critical temperature. In this case, a discontinuous phase transition occurs \cite{Pitaevskii}.
However, for large values of the magnetic field ($\omega\gg\varepsilon$), the transition becomes
diffuse \cite{Rojas1,Rojas2} and only crossover temperatures can be defined. Specifically, due to the
presence of the $N_{k}$ term, two different crossover temperatures will be identified.

\begin{figure}[t!]
  \null\hfill\includegraphics[width=0.45\textwidth,height=0.28\textheight]{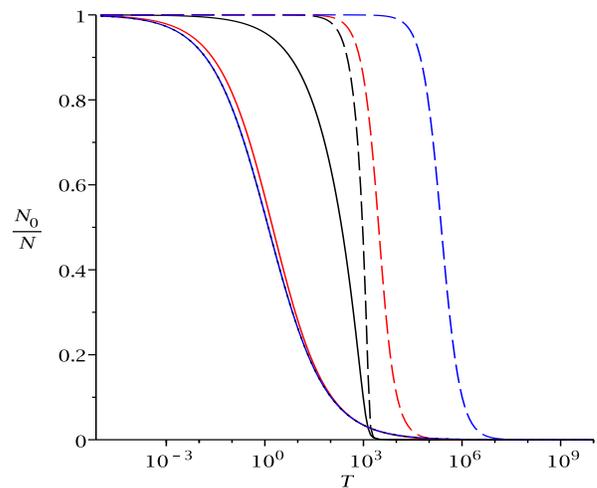}\hfill\null
  \caption{(Color online) Computed relative weight of $N_{0}$ for $N=10^{5}$ and $\omega=10^{2}$ (black lines),
 $\omega = 10^{5}$ (red -light gray- lines) and $\omega = 10^{8}$ (blue -gray- lines), in units of $\varepsilon$. 
Temperature is given in units of $\varepsilon/k_B$. The
conventional approach (which ignores $N_k$) is plotted with dashed lines. The effect of including the $N_{k}$ in the
relative weight of $N_{0}$ is plotted with solid lines. The limiting function $(1+\eta_{k})^{-1}$
is represented with black dotted lines (it overlaps with the solid blue (gray) line) [see Eq. (\ref{lim_N_O_N})].}
\label{fig1}
\end{figure}

It is quite remarkable that, for these high magnetic fields, $N_k$ gives
a significant contribution to $N$, as revealed in Fig.~\ref{fig1}, where
we show the relative weight of $N_{0}(T)$ described
by both the conventional and extended approaches ($N_{k}=0$ and $N_{k}\ne 0$, respectively)
for several values of the magnetic field.
From this figure it can be noted not only
that the phase transition becomes diffuse, but also that the macroscopic occupation of bosons with kinetic excited states
lying in the ground Landau manifold is very important at large magnetic fields.

We note that, within this limit of large magnetic fields and for $z\ll 1$ (i.e. $\beta|\mu|=-\beta\mu\gg 1$, high temperature limit for the effective 1D system), the temperature dependence of $N_0/N$ tends to
\begin{equation}\label{lim_N_O_N}
  \frac{N_0}{N} = \frac{1}{1+ \eta_{k}},
\end{equation}
as shown in Fig.~\ref{fig1}. In fact, this limit function cannot be distinguished from the case $\omega = 10^{8}$ depicted
in Fig.~\ref{fig1}.

\begin{figure}[t!]
  \null\hfill\includegraphics[width=0.45\textwidth,height=0.28\textheight]{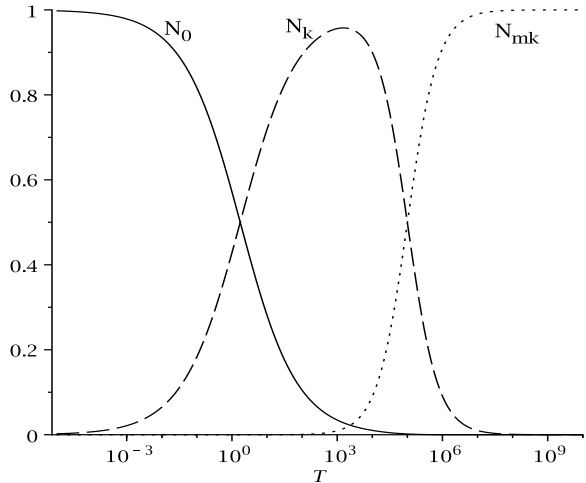}\hfill\null
  \caption{Computed relative weights of $N_{0}$ (solid line), $N_{k}$ (dashed line) and $N_{mk}$ (dotted line)
for $\omega=10^{5}$ and $N=10^{5}$. See text for details.}
\label{fig3}
\end{figure}

The origin of the discrepancy shown between the conventional and the extended approach (which increases
with the magnetic field) can be traced
to the relative weights of the different terms entering Eq. (\ref{analy_equation}), which explicitly includes
the contribution of $N_{k}$. Take for example
$\omega =10^{5}$ and $N=10^{5}$. In this case, as shown in Fig.~\ref{fig3},
$N_{k}$ represents a significant (and occasionally dominant) amount of the total particle
number. Moreover, two different crossover
temperatures appear when either $N_{k}$ or $N_{0}$ begin to grow substantially. In particular,
the phase transition that occurs at a lower temperature remains somewhat more diffuse than the high--temperature
one, because of its effective 1D nature.

\section{Specific heat}

%

\begin{figure}[t!]
  \null\hfill\includegraphics[width=0.45\textwidth,height=0.28\textheight]{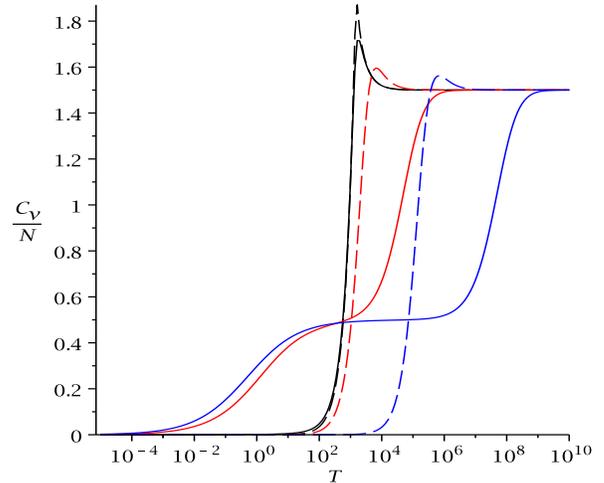}\hfill\null
  \caption{(Color online) Specific heat at constant volume as a function of temperature. The extended
and conventional approaches
are plotted with solid and dashed lines, respectively.
We have taken $N=10^{5}$ and $\omega=10^{2}$ (black lines), $\omega=10^{5}$ (red -light gray- lines)
 and $\omega= 10^{8}$ (blue -gray- lines). See text for details.}
\label{fig5}
\end{figure}

The essential features of BEC as a phase transition are clearly exhibited in the behavior of the specific heat. In particular,
we consider the specific heat at constant volume, defined by $C_{V}=-\beta^{2}\left(\partial U / \partial \beta\right)_{V}$.
In the extended approach, the internal energy can be expressed as
\begin{equation}
\label{int_ener}
U(\beta,z)= \frac{d}{2}\eta_{k} \beta^{-1} g_{3/2}(z)+
\frac{3d}{2}\eta_{k}\eta_{\omega} \beta^{-1} g_{5/2}(z).
\end{equation}
When no magnetic field is present, $C_{V}$ tends to the 3D
free Bose gas, displaying a sharp maximum and a discontinuous first derivative. The effect of a nonzero magnetic field is that of smoothing out that maximum \cite{Toms5}. This can be clearly appreciated in Fig. \ref{fig5}. Although the phase transition
remains diffuse in the presence of a magnetic
field, as previously discussed, one can still associate a critical temperature to the maximum of the specific heat,
as in the free gas case \cite{Toms5}. In Fig.~\ref{fig5} it can also be seen that, for $\omega=10^{2}$,
no significant deviations from the conventional
approach are found except for one: the inclusion of the $N_{k}$ terms tends to round off the relatively
sharp maximum shown in the specific heat calculated from the conventional approach.
On the contrary, for $\omega=10^{8}$,
significant differences between the conventional (dashed line) and extended (solid line) persist in a
wide range of temperatures.
Specifically, for large magnetic fields ($\omega\gg\varepsilon$) and
high temperatures ($\beta\omega\ll1$, or $\eta_m \gg 1$), $C_{V}/N$ approaches the
classical value of $3/2$. Within this regime, the classical limit can be obtained by taking
$N\simeq d \eta_{k}\eta_{\omega}z$
and $U\simeq\frac{3}{2}d\eta_{k}\eta_{\omega}\beta^{-1}z$, which results in $U=3N/2\beta$.
This is the classical limit for a system in three dimensions.

It is interesting to note that signatures of a double phase transition appear in the
specific heat. In particular, Fig.~\ref{fig5} shows
a plateau--like structure in the specific heat when the effective number of particles, $N/d$,
is of the order of $10^{-3}$ (solid blue -gray- line). This plateau spans the temperature range $\varepsilon \ll k_B T \ll \omega$. As $T$ decreases, the first phase transition, observed at a high temperature, reveals the
transition from states which have both kinetic and magnetic excitations to those which
are in excited kinetic states but in the ground Landau manifold. The second phase transition
shows how the absolute ground state (kinetic and magnetic) is reached. In fact,
the existence of two crossovers can be understood from Fig.~\ref{fig3}. This is
in accordance with the large slope of the $C_{V}(T)$ curve observed for the high--temperature
crossover, compared to the low--temperature one.
On the contrary, the specific
heat calculated within the conventional approach (that which ignores $N_k$) does not show signatures
of double phase transitions, as expected.

We remark that, in the plateau $\varepsilon \ll k_B T \ll \omega$, a constant value for the specific heat is obtained
between the two crossover temperatures. Within this temperature range,
$N\simeq d \eta_{k}z$ and $U\simeq \frac{1}{2}d\eta_{k} \beta^{-1}z$.
Finally, we get $U\simeq N/2\beta$. Thus, the plateau of the specific heat
lies in the value $C_{V}=Nk_{B}/2$, as shown in Fig.~\ref{fig5}, which
corresponds to that of a Bose gas in one dimension. We note that, although the value of the specific heat in the intermediate temperature plateau is independent of the degeneracy $d$, the details of the low and high temperature behavior depend on the particular value of $d$. In fact, we recall that the very existence of the plateau relies essentially on the degeneracy being much greater than unity.


\section{Conclusions}

In this work we have revisited the free charged Bose gas focusing on the case of a large magnetic field. Although a sharp Bose--Einstein
condensation does not occur, two diffuse phase transitions (or crossovers) are found. As the temperature is decreased, the first crossover reflects the enhanced occupation of states in the lowest Landau level. Then a range of intermediate temperature is found where the system behaves has an effectively one-dimensional system in its high-temperature regime. At even lower temperatures, the system undergoes a 1D Bose condensation, which is known to be a diffuse phase transition. To account for this two-step condensation, it has proved essential to include the occupation of states in the lowest Landau level but with arbitrary kinetic energy in the direction parallel to the magnetic field.

This work has been supported by MINECO (Spain)
through grants FIS2010-21372, CTQ2008-02578/BQU and the Juan de la Cierva program (PB).

\end{document}